\documentclass[twocolumn,showpacs,preprintnumbers,amsmath,amssymb]{revtex4}

\usepackage{graphicx}
\usepackage{dcolumn}
\usepackage{bm}
\usepackage[latin1]{inputenc}

\begin{document}

\title{Quantum localized modes in capacitively coupled Josephson junctions}
\author{R. A. Pinto and S. Flach}
\affiliation{Max-Planck-Institut f\"ur Physik komplexer Systeme, N\"othnitzer
Str. 38, 01187 Dresden, Germany}

\date{\today}

\begin{abstract}
We consider the quantum dynamics of excitations in a system of two
capacitively coupled Josephson junctions. 
Quantum breather states are found in the middle of the energy spectrum 
of the confined nonescaping states of the system.
They are characterized by a strong excitation of one junction.
These states perform
slow tunneling motion from one junction to the other,
while keeping their coherent nature. The tunneling time
sensitively depends on the initial excitation energy.
By using an external bias as a control parameter, the
tunneling time can be varied with respect to the
escape time and the experimentally limited coherence time.
Thus one can control the flow of quantum excitations between
the two junctions. 
\end{abstract}


\maketitle

\begin{figure}[!t]
\begin{center}
\includegraphics[width=1.5in]{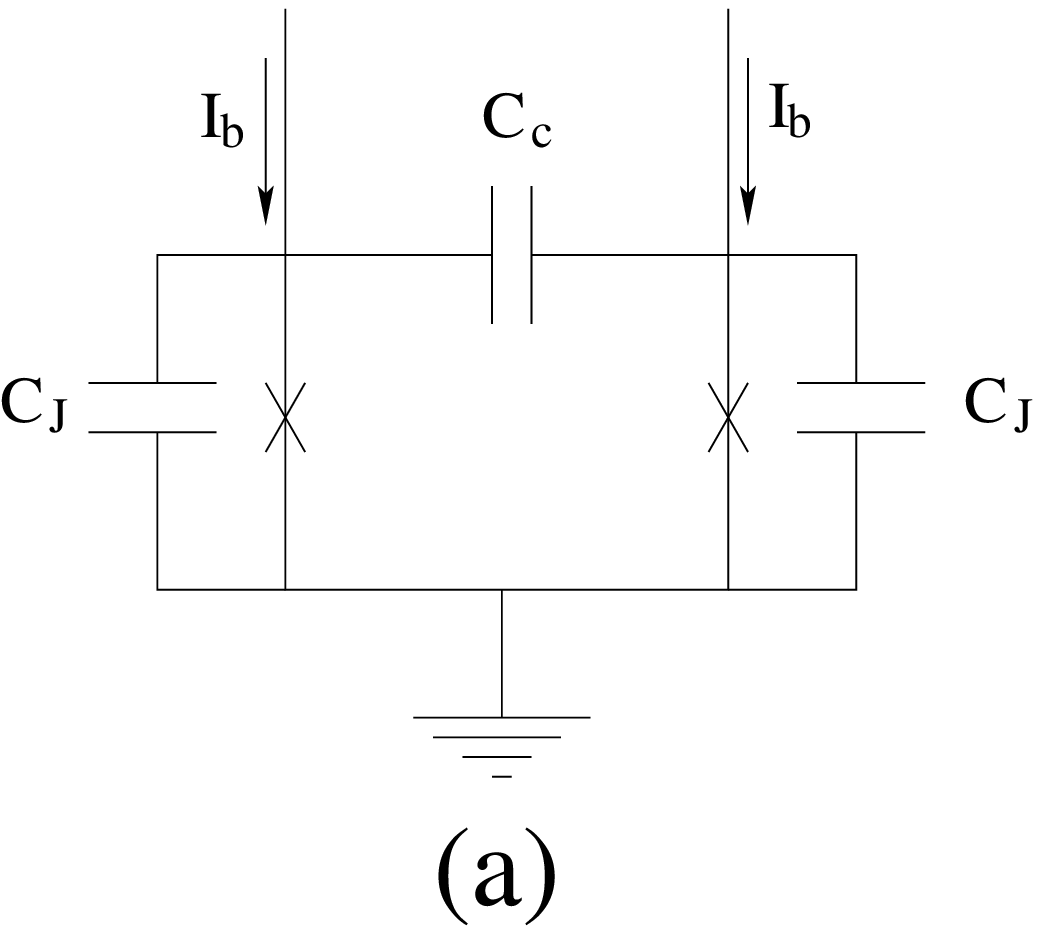}
\hspace{0.7cm}
\includegraphics[width=0.8in]{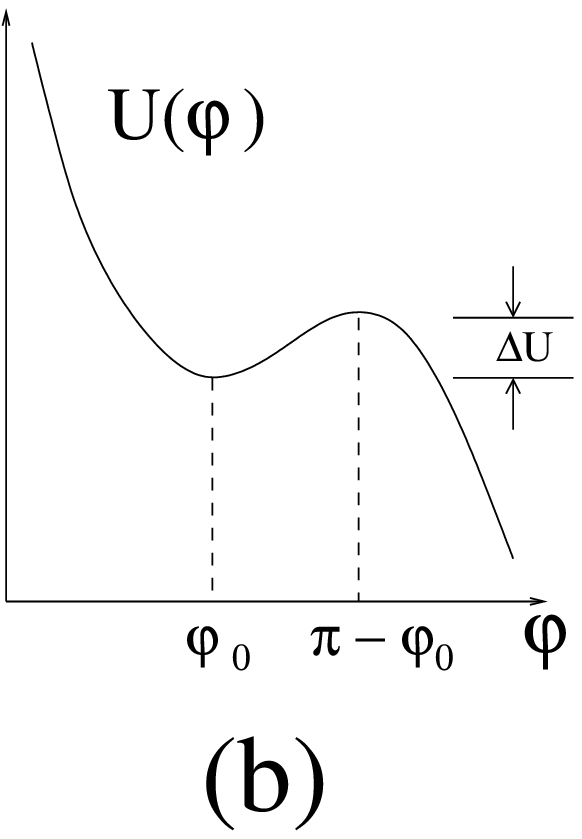}
\vspace{0.1cm}
\includegraphics[width=1.4in]{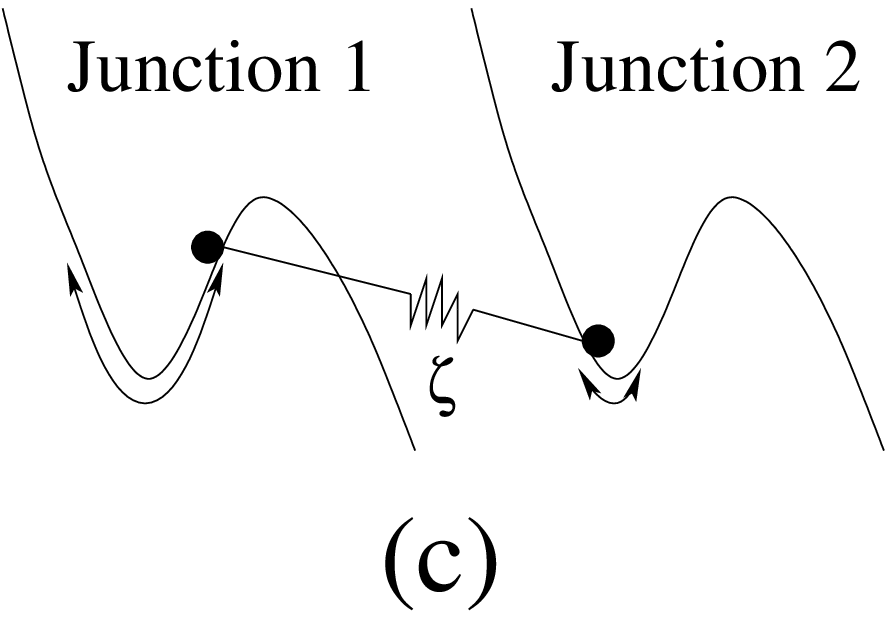}
\end{center}
\caption{\label{fig1}(a) Circuit diagram for two ideal capacitively coupled
JJs. (b) Sketch of the washboard potential for a single
current-biased JJ. (c) Sketch of a breather solution in the classical dynamics
of the system.}
\end{figure}

Josephson junctions are the subject of extensive studies in
quantum information experiments because they possess two attractive properties:
they are nonlinear devices,
and also show macroscopic quantum behavior \cite{Likharev,Leggett,Esteve}.
The dynamics of a biased
Josephson junction (JJ) is analogous to the dynamics of a particle with a mass
proportional to the junction capacitance $C_J$,
moving on a tilted washboard potential
\begin{equation}
U(\varphi) = -I_c\frac{\Phi_0}{2\pi}\cos\varphi - I_b\varphi\frac{\Phi_0}{2\pi},
\end{equation}
which is sketched in Fig.\ref{fig1}-b.
Here $\varphi$ is the phase difference between the macroscopic wave functions
in both superconducting electrodes of the
junction, $I_c$ is the critical current of the junction, and $\Phi_0=h/2e$ the
flux quantum. When the energy of the particle is large enough to overcome the barrier
$\Delta U$ (that depends on the bias current $I_b$) it escapes and moves down the
potential, switching the junction into a resistive state with a nonzero voltage
proportional to $\dot{\varphi}$.
Quantization of the system leads to discrete energy levels inside the wells in
the potential, which are nonequidistant because of the anharmonicity. Note that even if there is not
enough energy to classically overcome the barrier, the
particle may perform a quantum escape and tunnel outside the well, thus switching the junction into 
the resistive
state \cite{Likharev}. Thus each state inside the well is characterized by a bias and a state-dependent 
inverse lifetime, or escape rate.

Progress on manipulation of quantum JJs includes
spectroscopic analysis, better isolation schemes, and simultaneous measurement techniques
\cite{Leggett,Esteve,Steffen2006,Martinis,Steffen,Martinis2}, and
paves the way for using them as JJ qubits
in arrays for experiments on processing quantum information.
Typically the first two or three quantum levels of one
junction are used as quantum bits. Since the levels are nonequidistant, they
can be separately excited by applying microwave pulses.

So far, the studies on JJ qubits focused
on low energy excitations involving the first few energy levels of the
junctions. Larger energies
in the quantum dynamics of JJs
give rise to new phenomena that can be observed by using the already
developed techniques for quantum information experiments. For instance, it was suggested that Josephson
junctions operating at higher energies may be used for experiments on quantum chaos
\cite{Graham,Montangero,Pozzo}.
Another phenomenon is the excitation of quantum breathers (QB) \cite{Fleurov,Scott,Wang}, which are
nearly degenerate
many-quanta bound states in anharmonic lattices. When such states are excited, the outcome is
a spacially localized excitation with a very long time to tunnel from one lattice site to
another. So far the direct observation of this kind of excitations evolving in
time has not been reported. Up to date evidences of QB excitations were
obtained spectroscopically in molecules and solids
\cite{Fillaux1990,Fillaux1998,Richter1988,GuyotSionnest1991,Dai1994,Chin1995,Jakob1996,
JakobPr75,Okuyama2001,Edler2004}.

In this work we consider large energy excitations in
a system of
two capacitively coupled JJs \cite{Berkley,Johnson,Blais}. 
We study
the time evolution of states when initially only one of the 
junctions was excited. In the low energy sector such a state will lead to
a beating with a beating time depending solely on the strength of the
capacitive coupling.
For larger excitation energies the states perform slow tunneling motion,
where the tunneling time sensitively depends on the initial energy, in contrast
to the low energy beating time.
We calculate the eigenstates and the spectrum of the system, and identify quantum breather
states as weakly splitted tunneling pairs of states
\cite{Flach1,Pinto}. 
These eigenstates appear in the middle of the energy spectrum
of the system and are characterized by correlations between the two junctions -
if one of them is strongly excited, the other one is not, and vice versa.
By exciting one of the junctions to a large energy, we strongly overlap with
QB tunneling states. Consequently we trap the excitation on the initially
excited junction on a time scale which sensitively depends on the amount of energy excited,
and on the applied bias. We describe how this trapping could be experimentally
observed in time using the nowadays used techniques for manipulating JJ qubits.

The system is sketched in Fig.\ref{fig1}-a: two JJs are coupled by a
capacitance $C_c$, and they are biased by the same current $I_b$. The strength
of the coupling due to the capacitor is $\zeta = C_c/(C_c+C_J)$.
The Hamiltonian of the system is
\begin{equation}
H = \frac{P_1^2}{2m}+ \frac{P_2^2}{2m} + U(\varphi_1)+U(\varphi_2) + \frac{\zeta}{m}P_1P_2,
\end{equation}
where
\begin{eqnarray}
m &=& C_J(1+\zeta)\left(\frac{\Phi_0}{2\pi}\right)^2 ,\\
P_{1,2} &=& (C_c+C_J)\left(\frac{\Phi_0}{2\pi}\right)^2 (\dot{\varphi}_{1,2} -
\zeta\dot{\varphi}_{2,1}).
\end{eqnarray}

When the junctions are in the superconducting state, they behave like
two coupled anharmonic oscillators. The plasma frequency is $\omega_p =
\sqrt{2\pi I_c/\Phi_0 C_J(1+\zeta)} [1-\gamma^2]^{1/4}$, and $\gamma =
I_b/I_c$ is the normalized bias current.
The classical equations of motion of the system admit
breather solutions \cite{FlachPhysRep295}, which are time periodic, and for which the energy is
localized predominantly on one
of the junctions (Fig.\ref{fig1}-c). These orbits are numerically computed with high accuracy by using Newton
algorithms \cite{Flach2,Aubry1}. At the studied energies the
classical phase space is mixed, and breathers are
located inside regular islands, which are embedded in a chaotic layer.
	
In the quantum case we compute the energy eigenvalues and the eigenstates of the system.
We neglect quantum escape for states which will not escape in the 
classical limit. Thus we use a simple model for the single
JJ, where the potential energy is changed by adding a hard wall
which prevents escape:
\begin{equation}
U_{q}(\varphi) = \left \{ \begin{array}{ll}
U(\varphi) & \textrm{if $\varphi \leq \pi -\varphi_0$} \\
\infty & \textrm{if $\varphi > \pi -\varphi_0$}
\end{array} \right.,
\end{equation}
where $\varphi_0 = \arcsin\gamma$ is the position of the minimum of the
potential and $\pi -\varphi_0$ gives the position of the first maximum to the right from
the equilibrium position $\varphi_0$ (Fig.\ref{fig1}-b).
We will later compare the obtained tunneling times with the true state dependent escape times.

The Hamiltonian of the two-junctions system is given by $\hat{H} = \hat{H}_1 +
\hat{H}_2 + \zeta \hat{V}$, where $\hat{H}_i = \hat{P}_i^2/2m +
U_q(\hat{\varphi}_i)$ is the single-junction Hamiltonian
and $\hat{V}=\hat{P}_1\hat{P}_2/m$ is the interaction that couples the junctions.
The eigenvalues $\varepsilon_{n_i}$ and eigenstates $|n_i\rangle$ of the
single-junction Hamiltonian $\hat{H}_i$ were
computed by using the Fourier grid Hamiltonian method \cite{Fourier}. Note
that $|n_i\rangle$ is also an eigenstate of the number operator $\hat{n}_i$
with eigenvalue $n_i$. In the harmonic approximation $\hat{n}_i=\hat{a}_i^{\dagger}\hat{a}_i$,
where $\hat{a}_i^{\dagger}$ and $\hat{a}_i$ are
the bosonic creation and annihilation operators.
Since only states with energies below the classical escape energy (barrier) are taken into account,
the computed spectra have a finite upper bound.
The perturbation $\hat{V}$ does not conserve the total number of
quanta $n_1+n_2$, as seen from the dependence of the 
momentum operators on the bosonic creation and annihilation operators in the harmonic approximation:
$\hat{P}_{1,2}= (\Phi_0/2\pi)\sqrt{(1+\zeta)C_J\hbar\omega_p/2}
(\hat{a}_{1,2}-\hat{a}_{1,2}^{\dagger})/i$.

The Hamiltonian matrix is written in the basis of product states of the
single-junction problem $\{|n_1,n_2\rangle = |n_1\rangle \otimes
|n_2\rangle\}$. The invariance of the Hamiltonian 
under permutation of the junction labels allows us to use symmetric and
antisymmetric basis states
\begin{equation}
|n_1,n_2\rangle_{S,A} = \frac{1}{\sqrt{2}}(|n_1,n_2\rangle \pm |n_2,n_1\rangle )
\end{equation}
to reduce the full Hamiltonian matrix to two smaller symmetric and antisymmetric 
decompositions of $\hat{H}$, which after diagonalization respectively give the symmetric and
antisymmetric eigenstates of the system.

In order to identify quantum breather states, whose
corresponding classical orbits are characterized by energy localization,
we define the correlation functions:
\begin{equation}
f_{\mu}(1,2) = \langle\hat{n}_1\hat{n}_2\rangle_{\mu}
\end{equation}
\begin{equation}
f_{\mu}(1,1) = \langle\hat{n}_1^2\rangle_{\mu},
\end{equation}
where $\langle\hat{A}\rangle_{\mu} =
\langle\chi_{\mu}|\hat{A}|\chi_{\mu}\rangle$,
$\{|\chi_{\mu}\rangle\}$ being the set of eigenstates of the system.
The ratio $0 \leq f_{\mu}(1,2)/f_{\mu}(1,1) \leq 1$ measures the site correlation of quanta:
it is small when
quanta are site-correlated (when there
are many quanta on one junction there are almost none on the other one) 
and close to one otherwise.
\begin{figure}
\includegraphics[width=3.3in]{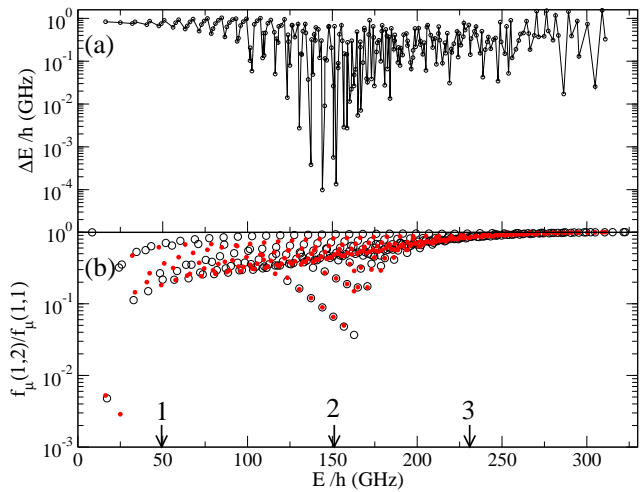}
\caption{\label{fig2}(a) Energy splitting and (b) correlation function
vs. energy for the two-junctions system (open circles, symmetric
eigenstates; filled circles, antisymmetric eigenstates). The labeled arrows mark the energy
corresponding to the peak of the spectral intensity in Fig.\ref{fig3}-b, d, and f (see text).
The parameters are
$\gamma=0.945$ and $\zeta =0.1$ (22 levels per junction).}
\end{figure} 

In Fig.\ref{fig2} we show the nearest neighbor energy spacing (splitting) and the correlation function of
the eigenstates. For this, and all the rest, we used
$I_c=13.3 \;\mu$A, $C_J=4.3$ pF, and $\zeta=0.1$, which are typical values in experiments.
We see that in the central part of the spectrum the energy splitting becomes
small in comparison to the average. The corresponding pairs of eigenstates, which are tunneling pairs,
are site correlated, and thus QBs. In these states
many quanta are localized on one junction and the tunneling time of
such an excitation from one junction to the other (given by the inverse energy
splitting between the eigenstates of the pair) can be exponentially large and depends
sensitively on the number of quanta excited.

The fact that the most site correlated eigenstates occur in the
central part of the energy spectrum may be easily explained as follows: Let $N$ be
the highest excited state in a single junction, with a corresponding
maximum energy $\Delta U$ (Fig.\ref{fig1}). For
two junctions the energy of the system with both junctions in the
$N$-th state is $2\Delta U$, which roughly is the width of the full
spectrum. Thus states of the form $|N,0\rangle$ and $|0,N\rangle$ that
have energy $\Delta U$ are located roughly in the middle.

\begin{figure}
\includegraphics[width=3.4in]{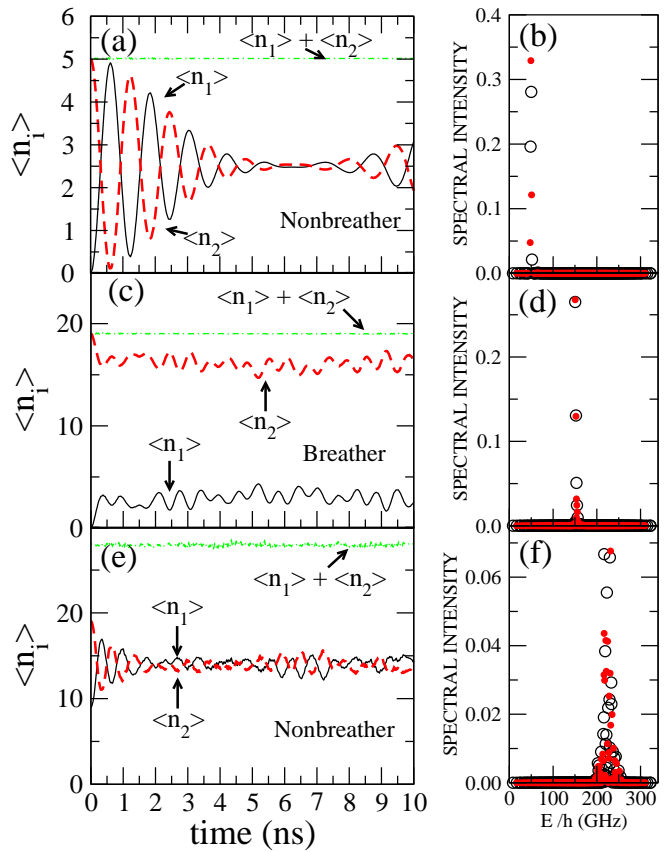}
\caption{\label{fig3}Time evolution of expectation values of the number of
quanta at each junction for different initial excitations and their
corresponding spectral intensity. (a) and (b): $|\Psi_0\rangle =|0,5\rangle$;
(c) and (d): $|\Psi_0\rangle =|0,19\rangle$;
(e) and (f): $|\Psi_0\rangle =|9,19\rangle$. Open circles, symmetric
eigenstates; filled circles, antisymmetric eigenstates. The energies of the peaks in the spectral
intensity are marked
by labeled arrows in Fig.\ref{fig2}-b (see text). The parameters are
$\gamma=0.945$ and $\zeta =0.1$ (22 levels per junction).}
\end{figure} 

With the eigenvalues and eigenstates we compute the time evolution of
different initially localized excitations, and the expectation value of
the number of quanta at each junction $\langle \hat{n}_i\rangle (t)=
\langle\Psi(t)|\hat{n}_i|\Psi(t)\rangle$. Results are shown in
Fig.\ref{fig3}a, c, and e. Also we compute the spectral intensity
$I_{\mu}^0 = |\langle \chi_{\mu}|\Psi_0\rangle|^2$, which
measures how strong the initial state $|\Psi_0\rangle$ overlaps with the
eigenstates.
Results are shown in Fig.\ref{fig3}-b,
d, and f, where we can see a peak in each case, which corresponds to the 
arrows in Fig.\ref{fig2}-b.
We can see that the initial state $|\Psi_0\rangle=|0,5\rangle$ overlaps with
site correlated eigenstates with an energy
splitting between them being relatively large and hence the
tunneling time of the initially localized excitation is short.
For the case $|\Psi_0\rangle=|0,19\rangle$ QBs are excited: The
excitation overlaps strongly with tunneling pairs of eigenstates in the
central part of the spectrum, which are
site correlated and nearly degenerate. The
tunneling time of such an excitation is very long, and thus keeps the quanta localized 
on their initial excitation site for corresponding
times.
Finally the initial state $|\Psi_0\rangle = |9,19\rangle$ overlaps with weakly site correlated
eigenstates with large energy
splitting. Hence the tunneling time is short.

We tested whether a (coherent or incoherent) spreading of the initial state over a suitable
energy window affects the results discussed above. For instance instead of using the basis state $|0,19\rangle$
as the initial state (Fig.\ref{fig3}-c,d), we superposed the
basis states $|0,20\rangle$, $|0,19\rangle$, $|0,18\rangle$, and $|0,17\rangle$. We found that
the results qualitatively do not change.

The experimental observation of QBs may be possible using the scheme of
McDermott {\it et al} for simultaneous state measurement of
coupled Josephson phase qubits\cite{Martinis}, where by applying current
pulses the time evolution of the occupation probabilities in the qubits is measured. By
applying a first microwave
pulse on one of the junctions we excite it into a high energy
single-junction state with energy $\varepsilon_l$ and leave the other one in the ground state.
In this
way we have an initial state similar to the ones shown in Fig.\ref{fig3}.
After a variable time period we apply simultaneous current pulses
to the junctions to lower their energy barriers $\Delta U$ and
enhance the probability of tunneling outside the potential well. Then we test which
junction switches to the resistive state (detected by a measurable voltage accross it). By
repeating the measuring many
times we obtain the populations in the junctions as a function of the time
period between the initial
pulse and the simultaneous measuring pulses.

Let us discuss the so far neglected quantum escape. For that
we computed
$\tau_{escape}$ by using the semiclassical formula \cite{Landau}
\begin{equation}\label{eq:semiclass}
\tau_{escape}^{-1}(\varepsilon) = \frac{\omega(\varepsilon)}{2\pi}\exp\left\{-\frac{2}{\hbar}
\int_a^b p(\varphi) d\varphi\right\},
\end{equation}
where $a$ and $b$ are the turning points of the classical motion in the
reversed potential at $U(\varphi)=\varepsilon$,
$p(\varphi)=\sqrt{2[U(\varphi)-\varepsilon]}$, and $\omega (\varepsilon)/2\pi$ is
the frequency of the oscillations inside the initial well.
In table \ref{times} we show the escape time from different metastable states,
and we compare it with the tunneling time $\tau_{tunnel}$ of an initial excitation
$|\Psi(0)\rangle = |0,l\rangle$ between the two junctions, estimated from the
energy splitting of the (symmetric-antisymmetric) pair of eigenstates with the largest overlap with
the initial excitation. We see that for $l=19$, where we excite QBs, the escape time is long enough
for observing at least one tunneling exchange between the two junctions
before escaping to the resistive state. Note that the cases $l=18$ and 17
also excite QBs which would show even more tunneling exchanges before
escaping. The case $l=16$ does not excite QBs but eigenstates that, though
having small energy splitting, do not show strong site correlation of quanta
as in the previous cases. From these results we expect that escaping to the resistive state will not prevent from the
experimental observation of QB excitations.

Another phenomenon that was not taken into account in our quantum model is
decoherence. 
To be able to observe tunneling between the junctions the coherence time 
has to be longer than the shortest tunneling time between the junctions, which is on the
order of 1 ns in the cases shown in Fig.\ref{fig3}-a and e. In the experiment
shown in \cite{Berkley} using a few levels per junction they obtained a coherence time on
the same order. However, in the experiment in \cite{Martinis} the coherence
time was about 25 ns, and more recently in \cite{Steffen} the coherence time was
approximately 80 ns. We expect that further improvements in experiments
\cite{Steffen2006} will give us longer coherence times.

Note that the above coherence times are shorter than the tunneling times of QB excitations
(see table \ref{times}), hence decoherence is an effect that can not be
ignored if one wants to do a more realistic quantum description of the
system. When exciting a JJ to a high-energy state, relaxation (over
dephasing) is usually the main source of decoherence. We can make a crude
estimation of the corresponding relaxation time $T_1$ by
using $T_1 \simeq hQ/\varepsilon_l$ ($Q$ is the quality factor of
the junctions), which holds for a harmonic
potential \cite{Martinis1987,Esteve1986}. For $l=19,18$ and 17,
$\varepsilon_l/h$ is around 150 GHz (see Fig.\ref{fig2}-b). For the JJs used in
\cite{Steffen2006}, $Q$ is between 500 and 1000, which leads to a relaxation time
between 3 ns and 6 ns. It is much smaller than the tunneling time of the
QB excitations, therefore one would expect to see instead of
tunneling, a freezing of the QBs on one of the junctions before they decohere due to relaxation. 

One could obtain more feasible results by increasing the bias current in such a way
that there are less energy levels in the junctions. With this, exciting a QB would
need less energy, and the relaxation time becomes longer. The
tunneling time of that QB excitation is shorter, and might be even shorter
than the relaxation time, allowing one to observe
tunneling before relaxation. This possibility, and the
inclusion of decoherence in our model, are
issues that will be addressed in a future work.
\begin{table}
\caption{Escape times for metastable states in a single JJ
$\tau_{escape}$ estimated by formula (\ref{eq:semiclass}), and tunneling time
of the initial excitation$|\Psi(0)\rangle = |0,l\rangle$ between the two
junctions $\tau_{tunnel}$ estimated from energy splittings.}
\label{times}
\begin{center}
\begin{tabular}{|c|c|c|}
\hline
$l$ & $\tau_{tunnel}$ (ns) & $\tau_{escape}$ (ns) \\ \hline
20 & 348 & 42  \\ \hline
19 & $1.8\times 10^3$ & $3.5\times 10^3$  \\ \hline
18 & $10.16\times 10^3$ & $503.2\times 10^3$  \\ \hline
17 & $2.3\times 10^3$ & $71.2\times 10^6$  \\ \hline
16 & 366 & $1.62\times 10^9$  \\ \hline
\end{tabular}
\end{center}
\end{table}

In summary, we have studied the classical and quantum dynamics of high-energy localized
excitations in a system of two capacitively coupled JJs.
In the classical case the equations of motion admit time periodic localized
excitations (discrete breathers) which can be numerically computed. For the
quantum case we showed that excitation of one of the junctions to a high level
leaving the another junction in the ground state lead to a QB with long
tunneling time. 
This is possible because the excitation overlaps
strongly with tunneling-pair eigenstates which live in the central part of the
energy spectrum and localize energy on one of the junctions. This result would not
qualitatively change if we excite a (coherent or incoherent) superposition of several product basis
states instead of only one.  We showed that with the available
techniques for manipulating JJ qubits the experimental observation of
QB excitations is possible. Escaping to the resistive state of the junctions
(which together with decoherence was not taken into account in our quantum model) would not
prevent us from doing that, and we expect that improvements in preparation
(higher quality factors) and isolation techniques of 
JJ will lead to long enough coherence times, such that
the phenomena we described in this work will be clearly observed.
That would ultimately pave the way of a controlled stirring of quanta on networks
of JJs.

\acknowledgments

This work was supported by the DFG (grant No. FL200/8) and by the ESF network-programme AQDJJ.

\end{document}